\documentclass[12pt]{article}

\usepackage{tocloft}
\usepackage[utf8]{inputenc}
\usepackage{csquotes}
\usepackage{sbc-template}
\usepackage{float}
\usepackage[inline]{enumitem}
\makeatletter
\newcommand{\keywords}[2][Keywords.]{%
      {\list{}{
      \relax
      \leftmargin=0.8cm
      \labelwidth=\z@
      \listparindent=\z@
      \setlength\topsep{-20 pt}
      \itemindent\listparindent
      \rightmargin\leftmargin}%
      \item[\hskip\labelsep                                    \bfseries\itshape #1]\itshape#2%
      \endlist}}
\makeatother
\usepackage[caption=false]{subfig}
\usepackage[dvipdfm]{graphicx}
\usepackage{url}
\usepackage[publish]{coop-writing}

\usepackage{tikz}
\usetikzlibrary{positioning, arrows}
\usetikzlibrary{shapes.misc}
\usetikzlibrary{shapes.arrows}
\usetikzlibrary{calc}
\usetikzlibrary{arrows,chains}
\usetikzlibrary{arrows.meta}
\usetikzlibrary{shapes.misc}
\usetikzlibrary{shapes.arrows}

\cwnamedef{xexeo}{red}{Xexéo}
\cwnamedef{marcus}{blue}{Marcus}

\cwdefanoncitetext{[OMITIDO, data]}
\cwdefanontext{[BLIND REVIEW]}

\cwblind{\EEEE}{4E}{[Model's Name]}

\usepackage{booktabs}
\hypersetup{colorlinks=false}
\hypersetup{pdfborder = {0 0 0}}

\title{A Conceptual Model for the Analysis of Investigation Elements in Games}

\author{\cwanon{Pedro Marques\inst{1} }, \cwanon{Marcus Parreiras\inst{1} }, \cwanon{Joshua Kritz\inst{1}}, \cwanon{Geraldo Xexéo\inst{1}} }

\address{
\cwanon{LUDES - Programa de Engenharia de Sistemas e Computação} \\
\cwanon{COPPE - Universidade Federal do Rio de Janeiro } \\ 
\cwanon{Av. Horácio Macedo, 2030, CT, Bloco H, sala 319, Rio de Janeiro, RJ - Brasil}
\email{\cwanon{\{pedromn, mparreiras,xexeo\}@cos.ufrj.br, joshuakritz2@gmail.com}}
}

\begin{document} 

\maketitle

\begin{abstract}

This paper presents the 4E conceptual model, developed to formally analyze investigation games from a game design perspective. The model encompasses four components: Exploration, Elicitation, Experimentation, and Evaluation. Grounded Theory was employed as the methodology for constructing the model, allowing for an in-depth understanding of the underlying concepts. The resulting model was then compared to existing literature, and its contributions were thoroughly discussed. Overall, the 4E model presents a comprehensive framework for understanding investigation games elements. It's application in two real-world scenarios demonstrates its practical relevance.

\end{abstract}


\section{Introduction} \label{sec:introducao}

This work presents a conceptual model for the formal analysis of the representation of the investigative process in games, named 4E, using game design elements as a viewpoint.

The investigative process is represented in various media and is often associated with the suspense and mystery genres. Classic examples are the works of Arthur Conan Doyle and Agatha Christie in literature, also adapted for audiovisual and video games~\cite{cawelti1976}.

The peculiarity of games as a medium lies in their power to represent processes and the interactivity they provide users \cite{bogost2010persuasive}.
Not assuming the role of Watson (the observer) but rather that of Miss Marple (the detective), the player embodies this dichotomy. 
This highlights the potential of games to grant agency to the player in the investigation process, effectively simulating an experience as an investigator. 
At the opposite end of this spectrum, we find the reader, instead of the player, following a predetermined investigative storyline, without exerting actual agency over its course.

There are various ways to model the investigation process, some closer to reality and others with a higher level of abstraction that focus on key aspects. 
By using such a model of the process, we can examine multiple investigations-related games and analyze their representation of the process.

Based on this context, this study aims to answer the following research question: \textbf{how game design elements related to investigation are structured in games?} This question  seeks to identify  game design patterns~\cite{Bjork2004}. By answering this question, we hope to gain valuable insights into ways of analyzing investigation in games and eventually formulate facilitative tools for their creation.

This work will be divided into sections, each addressing important aspects for the proposal of a framework for analyzing investigation games: Section \ref{sec:methodology} details the methodology used; Section \ref{sec:results} presents the resulting model (4E); Section \ref{sec:aplicacao} describes applications of the resulting model; Section \ref{sec:related} explains how the resulting model relates to the literature; finally, in Section \ref{sec:conclusao}, the conclusion of the work is presented, with comments on future work.

\section{Methodology}\label{sec:methodology}

We based our methodology on the principles of Grounded Theory (GT). 
A qualitative research method developed by Glaser and Strauss in the 1960s, GT enables the inductive development of theory from systematic data collection and analysis~\cite{Stol2016,hook2015}. 

GT can be broadly described as a process in which the researcher identifies the substantive area, collects data, performs open coding as data is collected, registers the process, moves to selective coding when categories are more stable, finds the theoretical codes to organize the substantive codes best, formulates a theory and finally reviews the literature to compare with the theory~\cite{hook2015}. 

Note that in GT ``...analysis of data is not a distinct phase. Data is gathered, analyzed and the results used to guide further data collection. This iterative process is ongoing, with data being gathered strategically as the process develops rather than according to a predetermined plan.''~\cite{hook2015}.  

Moreover, GT typically foregoes an extensive literature review before data collection and coding to avoid biasing research with preconceived ideas~\cite{hook2015}.

In the context of GT, the question of when to stop data collection arises. The concept of saturation then comes to help. It refers to the point in the data collection and analysis process where additional data do not provide any new information or insight to the categories that are being developed~\cite{hook2015}.

The data source for this study is a series of recorded game sessions, most of which are played by one of the authors. Thirty-two games, listed in Table \ref{tab:games}, were analyzed, with an estimated total of 2700 minutes. Most games were not played to the end, but only until it was clear that its characteristics were registered. 

\begin{table}[htb]
    \centering
    \caption{List of games analyzed.}
    \label{tab:games}
    \scriptsize
    \begin{tabular}{ll}
    \toprule
        A Hand With Many Fingers & Observer \\ 
        Paradise Killer & Outer Wilds \\ 
        Ace Attorney: Phoenix Wright & Overboard! \\ 
        Broken Sword & Ace Attorney Investigations: Miles Edgeworth \\ 
        Contradiction & Return of The Obra Dinn \\ 
        Danganronpa: Trigger Happy Havoc & Sam \& Max \\ 
        Detective Grimoire & Sherlock Holmes: Consulting Detective \\ 
        Discworld Noir & Sherlock Holmes: Crimes and Punishment \\ 
        Frog Detective 2 & Telling Lies \\ 
        Gemini Rue & The Blackwell Legacy \\ 
        Ghost Trick: Phantom Detective & The Last Express \\ 
        Heaven's Vault & Professor Layton and the Curious Village \\ 
        Heavy Rain & The Case of the Golden Idol \\ 
        Jenny LeClue - Detectivú & The Vanishing of Ethan Carter \\ 
        Tangle Tower & The Wolf Among Us \\ 
        L.A. Noire & Unheard \\ \bottomrule
    \end{tabular}

\end{table}

The recorded sessions were carefully reviewed and subjected to a data coding process, in which notable aspects of gameplay were systematically tagged and categorized. These characteristics included but were not limited to game mechanics, narrative elements, and player interactions. 

Once the data were coded and categorized, we embarked on the iterative process of comparing and contrasting these categories, refining them, and searching for overarching themes and patterns. Through this continual examination and re-examination of the data, we developed a set of theoretical constructs that provided insight into the dynamics of the gaming experience.

Inspired by Jesse Schell's concept of ``lens'' in ``The Art of Game Design,'' the identified categories were framed as ``lenses'' through which the gaming experience could be viewed and analyzed~\cite{schell2008art}. 
This approach allowed us to explore different aspects and nuances of the game experience in-depth, highlighting their potential influences on overall player engagement and satisfaction.

In the subsequent sections of this article, we delve deeper into the findings that emerged from this methodological exercise, presenting the key themes and the resultant theoretical framework, along with illustrative comments derived from the game session recordings.

\section{Results} \label{sec:results}

We found that gameplay elements related to investigation can be separated into four main categories: Exploration, Elicitation, Experimentation, and Evaluation. 
Each category has specific and unique subcategories (or themes). Based on this grouping, the concepts are formalized and presented in Section \ref{sec_4e_componentes} of this article.

Furthermore, by studying the intrarelationships among the four categories, it was possible to describe a cyclic and iterative process that provides a high-level procedural representation of the investigation in games. Section \ref{sec_4e_ciclo} provides an explanation of the cyclic view of the categories. The resulting model was named 4E, referring to the first letter of each main category's name.

\subsection{Introducing 4E Categories} \label{sec_4e_componentes}

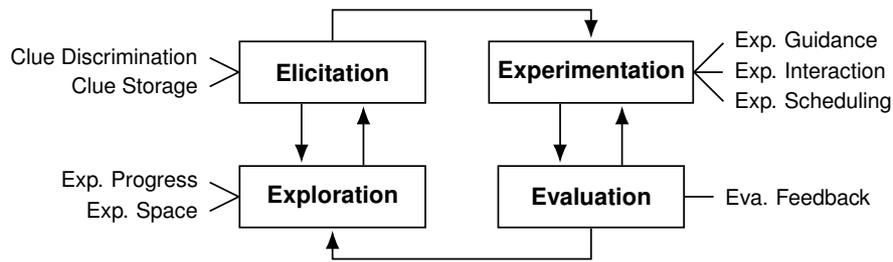
\begin{figure}[!ht]
\centering

\resizebox{0.8\textwidth}{!}{
\begin{tikzpicture}[
  ->,
  >=stealth',
  shorten >=1pt,
  auto,
  node distance=1cm,
  thick,
  main node/.style={
    rectangle,
    minimum width=3cm,
    minimum height = 1cm,
    draw,
    font=\sffamily\small\bfseries
  },
    sub node/.style={
    minimum height = 0.5cm,
    node distance=0.5cm and 0.5cm,
    font=\sffamily\footnotesize\mdseries
  }
]

  \node[main node] (1) {Elicitation};
  \node[main node] (2) [right=of 1] {Experimentation};
  \node[main node] (3) [below=of 2] {Evaluation};
  \node[main node] (4) [below=of 1] {Exploration};

  \draw[-{Latex[length=3mm]}] (1.north) 
  |- ($(1.north)+(0,.5)$) -| (2.north);
  \draw[-{Latex[length=3mm]}] (3.south) 
  |- ($(3.south)-(0,.5)$) -| (4.south);
  \draw[-{Latex[length=3mm]}] 
  ($(1.south)-(.5,0)$) -- ($(4.north)-(.5,0)$);
  \draw[-{Latex[length=3mm]}] 
  ($(2.south)-(.5,0)$) -- ($(3.north)-(.5,0)$);
    \draw[-{Latex[length=3mm]}] 
    ($(3.north)+(.5,0)$) --
  ($(2.south)+(.5,0)$);
  \draw[-{Latex[length=3mm]}] 
    ($(4.north)+(.5,0)$) --
  ($(1.south)+(.5,0)$);

    \node[sub node] (EF) [right=of 3] {Eva. Feedback};
   \draw[-] (3.east) -- (EF.west) ;

    \node[sub node] (XI) [right=of 2] {Exp. Interaction};
\node[sub node] (XG) [right=of 2,yshift=0.5cm] {Exp. Guidance};
\node[sub node] (XS) [right=of 2,yshift=-0.5cm] {Exp. Scheduling};
  
   \draw[-] (2.east) -- (XI.west) ;
   \draw[-] (2.east) -- (XG.west) ;
   \draw[-] (2.east) -- (XS.west) ;

\node[sub node] (CD) [left=of 1,yshift=0.25cm] {Clue Discrimination};

\node[sub node] (CS) [left=of 1,yshift=-0.25cm] {Clue Storage};

\node[sub node] (EP) [left=of 4,yshift=0.25cm] {Exp. Progress} ;

\node[sub node] (ES) [left=of 4,yshift=-0.25cm] {Exp. Space};

  \draw[-] (1.west) -- (CD.east) ;
  \draw[-] (1.west) -- (CS.east) ;
  \draw[-] (4.west) -- (EP.east) ;
  \draw[-] (4.west) -- (ES.east) ;

\end{tikzpicture}}

\caption{4E Diagram}
\label{fig:faji_components}
\end{figure}

The four categories and their respective themes can be used as a ``magnifying glass'' to establish a specific focus on the object of study. 
While the categories are interconnected and mutually exert influence on one another, their individual examination facilitates comprehensive analysis.
Through this specialization, categories assist in the identification of patterns and the grouping of approaches according to their themes. In Figure \ref{fig:faji_components}, next to each category, we can see their respective themes.

4E's categories and themes should not be directly applied to the game, but to it's elements. As a game can be have multiple elements, each of them can relate to a diffent category and theme.
For example, in terms of Exploration, the same game can have elements of linear progress and elements of non-linear progress.

\subsubsection{Exploration}

Exploration refers to the process of discovering and collecting information in the environment. 
It involves the player actively navigating through various areas, uncovering hidden locations, unveiling secrets, and encountering new elements within the game. 
It may be intuitive to think of exploration in a geographical sense, but the concept can also be applied to explore other diverse sources of information as databases, character dialogues, and so on.

Beachum (2013) \nocite{beachum2013outer} discusses different aspects of exploration in games and introduces a distinct category, Curiosity-Driven Exploration, which ``can be described as any situation in which someone chooses to explore her environment (real or virtual) with the primary objective of expanding her knowledge or understanding of it''. 
The Exploration process is driven by curiosity about the mystery~\cite{beachum2013outer}. 
Generally, it involves thoroughly examining the game world, interacting with objects, characters, and environments to gather relevant information and uncover hidden details. Garcia (2019) \nocite{vinicius_exploracao} lists ways the player can be prompted to explore out of curiosity.



We found that common activities of Exploration consist of searching for clues, evidence, and information to progress in the investigation and solve the game's mysteries or puzzles. Two themes were observed to have high diversity among the games:

\begin{itemize}
    \item \textbf{Exploration Progress}: represents reflecting on the freedom that the player has in the process of exploration and acquiring knowledge.
    \item \textbf{Exploration Space}: represents reflecting on the spaces that the player has access to visit and discover evidence.
\end{itemize}

\textbf{Exploration Progress} elements can be seen within a spectrum from linear to nonlinear exploration. 
Linearity refers to a predetermined progression through the space that causes clue encountering and information gathering to occur in a sequential manner. 
This is often used to create a specific narrative or story arc. 
On the other hand, non-linearity refers to freedom and flexibility to explore the space in different orders and manners. In this case, players have the ability to choose their own paths and tackle the investigation from multiple angles. 
The exploration in \textit{Professor Layton}'s games is closer to a linear experience, with predetermined objectives and cutscenes. On the other hand, the exploration in \textit{Unheard} is closer to a non-linear experience, with players having complete freedom to explore each conversation.

\textbf{Exploration Space} elements describe the type of environment explored and where the evidence is located. 
It does not necessarily have to be geographical like \textit{L.A. Noire} city or \textit{Tangle Tower}'s mansion: players may be exploring a database containing videos as in \textit{Her Story}, texts as in \textit{Heaven's Vault}, or audio recordings as in \textit{Unheard}. 
\subsubsection{Elicitation}

In the context of investigation, ``elicitation'' is used to refer to the act of extracting or obtaining information to progress or uncover critical details~\cite{FBI}
In this paper, we use this word to describe elements of game design related to the processes of perception, acquisition, and storage of clues. 
Through elicitation, players aim to extract important clues from the Exploration Spaces that can shed light on the investigation. It requires careful observation of verbal and nonverbal clues.

Our analysis revealed the following Elicitation themes:

\begin{itemize}
    \item \textbf{Clue Discrimination}: represents reflecting on how easily evidence is discriminated against other irrelevant elements of the game.
    \item \textbf{Clue Storage}: represents reflecting on how the game internally stores, or does not, the evidence acquired by the player.
\end{itemize}

Elements of \textbf{Clue Discrimination} influence the process of discerning clues in noisy information. 
This degree of separation is achieved through visual and game design choices to enhance the visibility and distinctiveness of important clues. 
By incorporating elements that introduce ambiguity and complexity, the game challenges players to discern genuine leads from misleading or inconsequential information. 
Some approaches to Clue Discrimination are: in \textit{Sherlock Holmes: Crimes and Punishment}, most clues are visually remarkable objects in the world; in \textit{Telling Lies} videos, it is not explicitly clear what is a clue and what is not relevant.

A typical \textbf{Clue Storage} element is representing clues as items that are organized and stored in an inventory, but there are other approaches related to not storing within the game interface, demanding external methods, such as tnote taking of important information outside of the game.
This additional layer of engagement requires players to actively participate in the investigative process beyond the confines of the game, enhancing immersion and fostering a deeper sense of involvement in solving the mysteries. Some approaches to Storage: in \textit{Phoenix Wright: Ace Attorney} and \textit{The Case of the Golden Idol}, clues are stored as items in an inventory; \textit{Outer Wilds}'s clues are stored as automatic notes; in \textit{Unheard}, clues are not stored by the game and the player must manage them.

\subsubsection{Experimentation}

Experimentation refers to the process of trying different approaches, methods, or actions to test hypotheses or uncover new leads during the course of the investigation. 
Experimentation often demands the player's active involvement in conducting various tests, simulations, critical thinking, problem solving skills, and a willingness to explore alternative paths or solutions. 
It adds elements of discovery, trial and error, and scientific inquiry to the gameplay experience, fostering a sense of engagement and agency in the player's role as an investigator.

Successful experimentation in investigation games should lead to important discoveries, breakthroughs, or the generation of new leads that advance the investigation. We found three themes corresponding to experimentation:

\begin{itemize}
    \item \textbf{Experiment Interaction}: represents reflecting on how much interaction the process of experimentation has;
    \item \textbf{Experiment Scheduling}: represents reflecting on how time controls the moments of experimentation;
    \item \textbf{Experiment Guidance}: reflects on how the creation of experiments is induced in terms of freedom to build your own questions and solutions about the mystery.
\end{itemize}

Elements related to \textbf{Experiment Interaction} manifest themselves in various ways, describing the mechanics of player intervention in experiment building. 
Some approaches consist of representing this process as minigames, as in \textit{Detective Grimoire: Secret of the Swamp} system for deduction. 
Games such as \textit{Telling Lies} and \textit{Her Story} choose to represent it in a more subtle way, in the form of database queries. 
There are even examples of approaches without interaction, as in \textit{Professor Layton}'s games, that use dialogs and cutscenes to represent deductions from the character.

\textbf{Experiment Scheduling} elements are related to the pace of the experimentation moments. 
Pre-determining these moments can support pivotal plot points or milestones within the investigation and can help the narrative of the game. 
On the other hand, there are instances where investigation games grant players the autonomy to decide when to conduct experiments. 
Some approaches to Scheduling: in \textit{The Case of the Golden Idol}, players experiment whenever they want using the collected information; in \textit{Phoenix Wright: Ace Attorney}, there are specific story moments in which the player must make deductions and experiments.

\textbf{Experiment Guidance} elements can vary from allowing players to create their own mistery questions, to presenting players with specific questions to address. 
Questions act as waypoints, directing players toward particular areas of interest. The game instills a sense of purpose and orientation in the player's experimentation by providing concrete questions to answer.
Games such as \textit{Return of the Obra Dinn} and \textit{The Case of the Golden Idol} exemplify well-defined mystery questions, demanding that players fill in the gaps with collected objects. 
Notable examples of low guidance include \textit{Her Story} and \textit{Telling Lies}, which provide minimal information on the mystery and enable players to formulate their own questions based on curiosity.

\subsubsection{Evaluation}

Evaluation is a crucial step in the deduction process. 
After players have conducted experiments or actions within the game world based on their hypotheses, they receive feedback from the game that helps them assess the effectiveness and validity of their experiments. 
This feedback can take various forms, such as dialogues with characters, notifications, visual cues, or changes in the game environment.
Feedback informs players about the consequences of their choices, helping them determine whether their experiments were successful or if they need to revise their approach.

Evaluation allows players to make sense of the information they have gathered, refining their deductions and theories as they progress in the game. It plays a pivotal role in the progression of the investigation game. For evaluation, we found only one theme:

\begin{itemize}
    \item \textbf{Evaluation Feedback}: represents how the game validates (or not) the player's experiments, whether by providing objective or subjective responses.
\end{itemize}

During the analysis, \textbf{Evaluation Feedback} emerged in various ways, ranging from explicit forms with clear visual and textual indications confirming an experiment's outcome, to more indirect and subtle forms that leave room for interpretation and doubt. 
The first type provides a sense of validation, reinforcing the players' deductions. 
The second introduces subjectivity and doubt, adding complexity to the evaluation process.
Some approaches to Feedback are: in \textit{The Case of the Golden Idol} and \textit{Return of the Obra Dinn}, players don't receive feedback from individual experiments, but when they complete correctly many of them; In \textit{Tangle Tower}, every experiment end with clear validation; In \textit{Telling Lies}, the game has almost none explicit validation, leaving the player only with theories.
\subsection{4E as a Abstract Gameplay Loop} \label{sec_4e_ciclo}

Gameplay is the word of the game industry to describe ``the process of the player playing the game or encountering a challenge''~\cite{guardiola2016gameplay}. There are different formal definitions of gameplay and the gameplay loop, but in this article we focus on the overall description of gameplay using verbs~\cite{guardiola2016gameplay}.

The 4E model proposes that in investigation games, gameplay consists of iterative cycles of Exploration, Elicitation, Experimentation, and Evaluation. Imagine Figure \ref{fig:faji_components} as a pathway, the player starts with the Exploration component and can navigate following the arrows between the components.

In the main cycle, the player begins by \textbf{Exploring} the game space in search of clues.
Then, in the \textbf{Elicitation} process, clues appear and are recognized.
Based on one or more clues, the player can eventually \textbf{Experiment} and form hypotheses, which are then \textbf{Evaluated} during gameplay.

There are also two smaller cycles in Figure \ref{fig:faji_components}, called the Observation Cycle and the Deduction Cycle: the first is composed of the sequence Exploration and Elicitation; the second consists of the sequence Experiment and Evaluation, making deductions.

The end of the game mystery occurs with a final step in Evaluation. The expected outcome must be related to a correct deduction about the investigation, thus concluding the mystery.
\section{Evaluating 4E} \label{sec:aplicacao}

In this section, the 4E model will be applied to identify and categorize the gameplay elements of two games: \textit{Clue}, a classic investigation board game, and \textit{Her Story}, an acclaimed digital game. \textit{Clue} was selected as a widely known paradigmatic board game representing the classic ``whodunit'' investigation. 
On the other hand, \textit{Her Story} is a successful independent game with an open end, i.e., there is not a final correct solution to the mystery, posing a challenge to its analysis. 

These two games represent interesting use cases of the model: 
as 4E was initially directed to videogames, \textit{Clue} as boardgame represent a analysis challenge; 
\textit{Her Story} is considered an edge case by the innovative mechanics and structure
\footnote{\textit{Her Story} received BAFTA's 2016 Game Innovation award}.

\subsection{Applying 4E in ``Clue''}

\textit{Clue} \nocite{clueManual} is a board game in which players assume the roles of characters enclosed in a mansion where an assassination took place, and they need to solve the murder. 
The crime is defined by three random cards which are sealed in an envelope when the game starts. They represent place, weapon, and culprit. The remaining possible cards are then distributed to the players and serve as clues. 
The players play by moving through the mansion, inquiring other players about the cards they have, to guess what cards are in the envelope.

Clue is a game that features non-linear \textbf{Exploration Progress}, where each player explores the mansion and interrogates freely. 
Information asymmetry motivates different paths among players. The \textbf{Exploration Space} is the mansion board which is divided into corridors and rooms; however, clues can only be found in rooms.

The elicitation process has clear \textbf{Clue Discrimination} since they are explicitly and concretely represented by the cards in the player's hands. The \textbf{Clue Storage} is done through the individual and private annotations of the players on a game sheet.

\textbf{Experimentation Guidance} occurs in the structure of the guesses, you always name a murderer, a weapon and a place. Experiments are divided into two types: questions and an accusation. 
The questions are a \textbf{ Scheduled Experimentation} because players can do them once per turn. 
The accusation is not scheduled and can occur at any moment; as soon as players think they know all three cards of the envelope, they can make an accusation, however, if wrong, they are eliminated from the game. 
The \textbf{Experiment Interaction} is the player's choice of which cards to guess. However, the place guessed is where he is, and his agency allows only for choosing the weapon and the murderer.

The \textbf{Evaluation Feedback} of the guesses is given according to their type: for CLUE questions, the game always validates one card for the player who asked the question; for accusations, the game always reveals whether it was a correct or incorrect guess, leaving no doubts for the players.

\subsection{Applying 4E in ``Her Story''}

In \textit{Her Story}, the players search a database of police interview videos from a fictional investigation. 
The gameplay mechanics are minimalistic and involve searching for specific keywords within the video database and watching said videos. 
Players enter search terms into the game's interface, and the database retrieves videos containing those keywords. 
As players watch the videos, they gain insights into the fragmented and non-linear narrative. The game's mechanics encourage note-taking of their observations, connections, and theories as they progress through the videos. 

Her Story features a completely nonlinear \textbf{Exploration Progress}, where players are free to explore a database of video clips using keywords. 
There is no structure in the order the videos are found, and the gameplay sequence completely relies on player input. The \textbf{Exploration Space} in this game is represented by the video database, which contains information in the form of video clips.

The \textbf{Clue Discrimination} is a game process: the game does not clearly distinguishes information as relevant or irrelevant, and is up to the player to judge.
The \textbf{Clue Storage} is done through the player's individual note-taking or mental recollection of the information gathered from the video clips.
All video clips watched are stored in a video backlog, allowing players to return to previous media.

The \textbf{Experimentation Interaction} primarily involves the player's interactions with the game's interface, conducting searches and selecting video clips to watch, and making connections between the information obtained.
The \textbf{Experiment Schedule} theme revels that during the gameplay, there are no specific moments for creating experiments, so the pace is done by the players themselves as they formulate questions and become curious.
\textbf{Experimentation Guidance} in \textit{Her Story} only occurs in the first keyword in the game: ``Murder''. 
Other than that, players have the freedom to search for any keyword. 


The game does not give \textbf{Evaluation Feedback} about the player's actions but allows them to draw their own conclusions about the narrative based on the information they have gathered. 
The absence of feedback leaves room for different players to have varying understandings of the story. 
This game design decision brings the game closer to a more realistic investigative experience, encouraging players to question their own theories and to be more meticulous in taking notes about the videos.

\section{Related Work} \label{sec:related}

In the GT tradition, we now describe some related work and discuss the relationships with ours. This revision used generic search terms to investigate academic research on detective games \footnote{Currently, we are developing a game-by-game approach, aiming to find methods hidden in the analysis of specific games.} and literature.


Maybe this lack of more general studies on the Ludology of detective video games can be linked to Blye (2023) \nocite{Blye2023} statement that ``Video games revolving around being a detective are few and far between''. On the other hand, there is a large market for them, and some have achieved fame. L.A. Noire, an AAA game that sold around 5 million, got enthusiastic reviews~\cite{Meslow2017}. Games of lesser development cost,, such as ``Her Story'', and the ``Sam \& Max'' series, are considered a success. Hence, while our model leans more towards the Ludology tradition, the references we found draw from the Narratology tradition. 

However, it is worth noting that even within Literary Studies, there exists a subset of work that explores what one might term the ``mechanics'' of detective novels \cite{Ramos2021, Vara2018}. This suggests that a shift towards more mechanical or ludic analysis can indeed coexist within a primarily narrative-oriented field.

Ramos (2021) \nocite{Ramos2021} tries to explain how the literary conventions of the detective story format are transposed into video games. She appeals to literary studies to understand detective games and mainly uses Cawelti's Framework~\cite{cawelti1976} for classical detective story formula, which is composed of four main patterns: situation, action, characters and relationships, and setting.
Ramos's work aligns with the Narratology approach to analyzing video games. 
On the other hand, our work built a theory based on game elements, where the player is the agent, bringing the Ludology view into the scene~\cite{frasca1999}.
Cawalti's four patterns do not characterize how the game works, as our model does, but rather what the story is. 
We actually plan to investigate how those two models can be put together.

Fernandez-Vara (2018) \nocite{Vara2018}, also starting with a Narratology approach, identifies the  ``relationship between game design and narrative at two basic levels: structure and behavioral scripts.''. 
She points to Todorov (1977) \nocite{todorov1977}, which defines the dual nature of detective stories, the story of the crime, and the story of the investigation. 
This further divides detective fiction into ``whodunits'', where the main narrative is about investigations and puzzle solving, and ``thrillers'', which are about the adventure of the investigation. 
Our work focuses on games where investigation and puzzle solving are the main mechanics, i.e., ``whodunits'' games.

We also found out that our model  aligns with the ideas of \cite{guardiola2016gameplay}, which proposes to analyze games through the gameplay loop, which are described by diagrams that resemble state machines where each state is an action, and those actions can appear in different levels of aggregation. Both approaches focus on mechanics over narrative, use actions to model high-level interactions, and allow specifying them in more detailed actions. 
However, inspired by Schell (2008) \nocite{schell2008art}, we decided to codify the basic knowledge of our theory as open questions that allow an analyst or designer to choose specific answers for each game.




\section{Conclusion} \label{sec:conclusao}

This article presented a conceptual model for the formal analysis of the representation of investigative process in games. By using game design elements as a viewpoint, the model represents a tool to examine and analyze how investigations are structured within games. With this model now researchers and designers can scrutinize the elements of a detective game that define its investigative process. Which can be used to both improve and criticize the game.

As far as we know, ours is the first work that studied a reasonable amount of games to propose a theory for what can be described as a Ludology approach to discuss detective games. After the model evaluation, 4E was shown to be useful in capturing key elements of investigation games, both in videogames and boardgames. 


As future work, we believe that the 4E could be applied and evaluated in more games, particularly in investigative board games, which were not included in the game database. In addition, we believe that the model could support the creation of tools for game design, such as canvas or development methodologies. Finally, the categories of the 4E should be studied in relation to other narrative-based models, aiming to establish a connection between game dynamics and the narrative.





\bibliographystyle{sbc}
\bibliography{sbc-template}

\end{document}